\documentclass[10pt,prl,aps,twocolumn,preprintnumbers,superscriptaddress]{revtex4-2}

\usepackage{amssymb}
\usepackage{amsmath}
\usepackage{mathtools}
\usepackage[colorlinks=true, pdfstartview=FitV, linkcolor=blue, citecolor=blue, urlcolor=blue]{hyperref}
\usepackage{bm}
\usepackage{braket}
\usepackage{graphicx}
\usepackage{comment}
\usepackage[normalem]{ulem}

\newcommand{\rmi}{\mathrm{i}}
\newcommand{\rme}{\mathrm{e}}
\newcommand{\rmd}{\mathrm{d}}

\renewcommand\i{\mathrm{i}}

\newcommand\sect[1]{{\it #1}---}

\begin{document}

\title{Quantum-Geometric Meissner Effect in Magnetized Color Superconductors}

\author{Kazuya Mameda}
\affiliation{Department of Physics, Tokyo University of Science, Tokyo 162-8601, Japan}
\affiliation{RIKEN iTHEMS, RIKEN, Wako, Saitama 351-0198, Japan}

\author{Noriyuki Sogabe}
\affiliation{Department of Physics, The University of Osaka, Toyonaka, Osaka 560-0043, Japan}

\preprint{RIKEN-iTHEMS-Report-26}

\begin{abstract}
We find a quantum-geometric Meissner response in magnetized two-flavor color-superconducting (2SC) quark matter.
Landau quantization quenches the transverse quasiparticle dispersion, suppressing the conventional Fermi-surface contribution and giving rise to a Meissner response governed by the quantum geometry of the Landau levels. In the strong-field regime, the response becomes dominated by the quantum metric of the lowest Landau level (LLL), and the leading scaling of the transverse Meissner mass is consequently set by the pairing gap, in contrast to the chemical potential scaling of conventional color superconductors.
This unconventional scaling has a topological origin, as the LLL quantum metric is constrained by its Chern number.
The reduced transverse Meissner mass brings potential implications for kHz quasi-periodic oscillations in magnetars.
\end{abstract}

\maketitle

\sect{Introduction}%
Quantum geometry is a ubiquitous concept across modern quantum physics~\cite{Liu_2024,Gao2025,Verma,Jiang_2025,yu2025quantum}. 
Alongside its well-known applications in transport, quantum geometry plays a key role in superconductivity~\cite{Kopnin2011,Peotta2015}.
A conventional Meissner response is dictated by the quasiparticle Fermi velocity, and is thus naively expected to vanish in dispersionless flat bands. 
However, even when the energy dispersion is quenched, the quantum states themselves can vary in momentum space, with their variation characterized by the quantum metric.
The resulting quantum-geometric contribution to the Meissner response explains the robustness of superconductivity in flat-band systems, e.g., twisted bilayer graphene~\cite{PhysRevLett.123.237002,PhysRevB.101.060505,Torma_2022,Torma:2023tqu}.

In this Letter, we demonstrate that quantum geometry also plays an essential role in color superconductivity in magnetized dense quark matter~\cite{Ferrer:2005vd,Fukushima:2007fc,Noronha:2007wg,Alford:2007xm}.
In such a phase, the electromagnetic and color magnetic fields are mixed to form the rotated magnetic field $\tilde B$ and the orthogonal field $B_X$. The former is not expelled, whereas the latter is subject to the Meissner effect~\cite{Alford:1999pb}.
In the presence of $\tilde B$, the two-flavor color-superconducting (2SC) matter provides a natural relativistic realization of a flat-band superconductor:  
Landau quantization quenches the transverse dispersion of the quasiparticles. 
We show that the transverse Meissner mass $M_\text{2SC}$ in the strong-field regime is dominated entirely by the quantum metric of the lowest Landau level (LLL), with the leading scaling set by the pairing gap $\Delta$:
\begin{equation}
\label{eq:M2SC}
M_\text{2SC} \sim |\Delta|\,.
\end{equation}

The gap scaling in Eq.~\eqref{eq:M2SC} reflects the suppression of the conventional Fermi-surface scale,
$M_\mathrm{conv} \sim \mu$ governed by the quark chemical potential $\mu$~\cite{Bailin:1983bm,Son:1999cm,Rischke:2000qz,Rischke:2000ra}, due to Landau quantization.
The conventional scaling persists even in the magnetized color-flavor-locked (CFL) phase, as $\tilde{B}$-neutral pairing channels evade Landau quantization~\cite{Sen:2015cxa}.

Nevertheless, no magnetic-field scale appears in the leading scaling of the Meissner mass.
We show that the absence of this magnetic-field scale has a topological origin in the Landau levels.
In the strong-field regime, the LLL quantum metric saturates the topological bound set by the Chern number and scales inversely with the Landau degeneracy, thereby canceling its magnetic-field factor.
Remarkably, the Meissner response becomes dominated by the topologically constrained LLL contribution in the strong-field regime, a feature absent in Bloch-band systems such as the Harper–Hofstadter model, where the continuum Landau-level regime is realized only in the weak-field limit~\cite{Peotta2015,PhysRevB.104.045103}.

Magnetars provide a possible astrophysical environment for magnetized color superconductors.
Although smaller than the conventional Meissner scale, the reduced quantum-geometric Meissner mass may provide a characteristic signature of strongly magnetized 2SC matter in their cores.
We show that Eq.~\eqref{eq:M2SC} introduces an intermediate propagation scale for magnetic disturbances in magnetars, between relativistic color-superconducting core dynamics and crustal Alfv\'enic modes.
This scale may be relevant for the kHz quasi-periodic oscillations (QPOs) observed in magnetar flare tails.

\sect{Quantum Geometry of Landau-quantized Weyl fermions}%
To analyze the quantum-geometric effect in the magnetized 2SC phase, 
let us first consider the general quantum geometry of Weyl fermions.
The right-handed Weyl Hamiltonian $H$ under an external rotated magnetic field $\tilde{\bm{B}}=\tilde{B}\hat{z}$ is given by $H = \bm{\sigma}\cdot(-\i\bm{\nabla} - \tilde{q}\tilde{{\bm A}})$, where $\bm{\nabla}\times{\tilde{\bm{A}}}=\tilde{\bm{B}}$ and $\sigma_i$ ($i=x,y,z$) denote the spinor-space Pauli matrices.
The eigenstates are defined as $H|u_N\rangle=\varepsilon_N|u_N\rangle$, and labeled by $N=\{n,\lambda\}$ with the Landau-level index $n$ and the energy branch $\lambda=\pm$, together with the longitudinal momentum $k$:
\begin{align}
\label{eq:eigen_state}
&|u_N\rangle
= \sqrt{\frac{\varepsilon_N+k}{2\varepsilon_N}}
|n,k,\uparrow\rangle
+\lambda\sqrt{\frac{\varepsilon_N-k}{2\varepsilon_N}}
|n-1,k,\downarrow\rangle\,,
\end{align}
where $\uparrow, \downarrow$ label the eigenstates of $\sigma_z$.
The energy eigenvalue $\varepsilon_N$ is given by $\varepsilon_N=k$ for $n=0$ and $\varepsilon_N = \lambda\sqrt{k^2+n\omega_{\tilde B}^2}$ for $n\geq 1$, where $\omega_{\tilde{B}}\coloneqq \sqrt{2\tilde{q}\tilde{B}}$.

When energy eigenstates are not labeled by conventional momentum variables, the quantum geometric tensor (QGT)~\cite{ProvostVallee1980,Shapere:1989kp} can be defined in terms of the position
operator~\cite{Bellissard,Marzari:1997obb,Bianco}.
We adopt this formulation for the present system under a magnetic field, and the QGT is defined in the 
gauge-invariant form as follows~\cite{Bellissard}:
\begin{equation}
(Q_{ij})_N
\coloneqq
-\langle u_N|
 [x_i^\perp,P_N][x_j^\perp,P_N] |u_N\rangle\,,
\label{eq:QGT}
\end{equation}
where $P_N=|u_N\rangle\langle u_N|$ and $\bm{x}^\perp = (x,y)$.
As usual, the quantum metric $g_N$ and Berry curvature $\Omega_N$ are defined as the real and imaginary parts of the QGT, respectively: $g_N\coloneqq \mathrm{Re} (Q_{xx})_N = \mathrm{Re} (Q_{yy})_N$ and
$\Omega_N= -2\mathrm{Im} (Q_{xy})_N=2\mathrm{Im} (Q_{yx})_N$.
Inserting the completeness relation and using the Landau-level selection rule, we reduce $g_N$ and $\Omega_N$ to
\begin{align}
g_N
&= \sum_{\lambda'}(a_{n,n+1}^{\lambda\lambda'}+a_{n,n-1}^{\lambda\lambda'})\,,
\label{eq:gN}\\
\Omega_N
&=2\sum_{\lambda'}(a_{n,n+1}^{\lambda\lambda'}-a_{n,n-1}^{\lambda\lambda'})\,,
\label{eq:OmegaN}
\end{align}
where $a_{NN'}\coloneqq|\langle u_N|x|u_{N'}\rangle|^2=|\langle u_N|y|u_{N'}\rangle|^2$.

Within the position-space formulation,
the Chern number is defined as the trace of the Berry curvature per unit area~\cite{Bellissard}.
In the present case, this trace reduces to the expectation value in the representative state $|u_N\rangle$ multiplied by the Landau-level degeneracy factor $\omega_{\tilde B}^2/(4\pi)$, yielding the Chern number independent of $N$:
\begin{equation}
\label{eq:Chern}
\begin{split}
 C
 &\coloneqq -2\pi\rmi 
 \frac{\omega_{\tilde B}^2}{4\pi}\langle u_N| \bigl[[x,P_N],[y,P_N]\bigr] |u_N\rangle
 = \frac{\omega_{\tilde B}^2\Omega_N}{2} 
 = 1\,,
\end{split}
\end{equation}
where the last equality follows from $\Omega_N=2/\omega_{\tilde B}^2$ for arbitrary $k$ and $N$.
Equations~\eqref{eq:gN}--\eqref{eq:Chern} lead to $g_N \ge |C|/\omega_{\tilde B}^2$, 
where the equality is attained only for the LLL ($n=0,\lambda = +$) since the downward transition is absent.

\sect{Spectral representation of Meissner mass}%
Let us now analyze the magnetized 2SC matter, where the superconducting quark pairs are $u_r$-$d_g$ and $u_g$-$d_r$.
We define $\widetilde{\mathrm{U}}(1)$ and $\mathrm{U}(1)_X$ as the gauge groups associated with $\tilde{B}$ and $B_X$, and the corresponding charges of $u$- and $d$-quarks are $(\tilde q,q_X)\coloneqq( \tilde{e}/2,\,e_X/(2\sqrt3))$ and $(-\tilde q,q_X)$, respectively~\cite{Alford:1999pb}.
In the weak photon-gluon mixing limit, we have
$\tilde e\simeq e$ and $e_X\simeq g$,
where $e$ and $g$ denote the electromagnetic and strong gauge couplings, respectively.

Writing the Nambu--Gor'kov (NG) spinor $\Psi=(u_r,\i\sigma_yd_g^*)^\mathrm{T}$,
we obtain the corresponding mean-field Bogoliubov--de Gennes action $S=\int \mathrm{d}^4x\,\Psi^\dagger\hat G^{-1}\Psi$.
In the imaginary-time formalism, the inverse NG propagator takes $\hat{G}^{-1}
=
-\partial_\tau-(H-\mu)\tau_3+\Delta \tau_+ +\Delta^* \tau_-$,
where $\tau_a$ ($a=1,2,3$) are the Pauli matrices in the NG space and $\tau_\pm=(\tau_1\pm \i\tau_2)/2$.
Here, we assume $\Delta$ is a uniform, momentum-independent pairing gap.

The transverse Meissner mass $M$ is obtained from the static quadratic response of the thermodynamic potential with respect to the $\mathrm{U}(1)_X$ gauge field.
This is expressed as $M^2=(\Pi_{xx}+\Pi_{yy})/2$, where $\Pi_{ij}$ is the total polarization tensor in the zero momentum and frequency limit. 
Due to the residual $\mathrm{SU}(2)_\mathrm{c}$ symmetry, the $u_g$-$d_r$ pair gives the same contribution as the $u_r$-$d_g$ pair considered above, 
while the left-handed sector gives an identical contribution, 
yielding an overall factor of four.
Hence, at the one-loop level, 
$\Pi_{ij} = q_\Delta^2\mathrm{Tr}\Bigl[\hat{G}\sigma_i\hat{G}\sigma_j\Bigr]$,
where $q_\Delta\coloneqq 2q_X$ denotes the $\mathrm{U}(1)_X$ charge of the Cooper pair.

While the conventional formulation evaluates the polarization tensor by performing the spin trace over quasiparticle propagators and vertices~\cite{Son:1999cm,Rischke:2000qz,Rischke:2000ra}, we instead employ a spectral representation that retains the transition matrix elements between Landau-level eigenstates explicitly, thereby revealing their relation to quantum geometry.
After inserting the complete eigenbasis of $\hat{G}^{-1}$ into $\Pi_{ij}$ and performing the Matsubara-frequency sum at zero temperature~\cite{supplemental}, we obtain
\begin{align}
\label{eq:M2_bare}
M^2(\tilde B)
=
\frac{q_\Delta^2\omega_{\tilde B}^{2}}{4\pi}
\sum_{N,N'}
\int \frac{\rmd k}{2\pi}
T_{NN'}(k)\,
S_{NN'}(k)\,,
\end{align}
where
\begin{align}
S_{NN'}
\coloneqq
\frac{
\xi_N\xi_{N'}
-
E_NE_{N'}
+
\Delta_\lambda
\Delta_{\lambda'}^*
}{
E_NE_{N'}(E_N+E_{N'})
}\,,
\label{eq:S}
\end{align}
with $\xi_N=\varepsilon_N-\mu$ and $E_N=\sqrt{\xi_N^2+|\Delta_{\lambda}|^2}$.
We adopt a simple constant-gap model, $\Delta_+ = \Delta$ and $\Delta_- =0$.
We also define the squared transition matrix element as
\begin{equation}
\label{eq:TNN'-def}
T_{NN'}\coloneqq\frac{1}{2} |\langle u_N|
\bm{\sigma}^\perp
|u_{N'}\rangle|^2\,,
\end{equation}
where $\bm{\sigma}^\perp = (\sigma_x,\sigma_y)$.
Evaluating $\langle u_N|\bm{\sigma}^\perp|u_{N'}\rangle$ with Eq.~\eqref{eq:eigen_state}, we obtain
$T_{NN'} =\mathcal{T}_{NN'}(k)\delta_{n', n+1}+ \mathcal{T}_{NN'}(-k)\delta_{n', n- 1}$ with
\begin{equation}
\begin{split}
\label{eq:Tnll'}
\mathcal{T}_{NN'}(k)
&=
\frac{(\varepsilon_N+ k)(\varepsilon_{N'}- k)}{4\varepsilon_N\varepsilon_{N'}}\,.
\end{split}
\end{equation}
Here, the selection rule $n'=n\pm1$ follows from the spin-flipping nature of the vertex $\bm{\sigma}^\perp$ and the spin-orbit locking encoded in the Landau-level eigenstates~\eqref{eq:eigen_state}.

As in the polarization tensors of other gauge theories, the Meissner
mass~\eqref{eq:M2_bare} contains a vacuum divergence arising from transitions between positive- and
negative-energy branches.
From the high-energy asymptotic expansion of $S_{NN'}$, we identify and subtract
$S_{NN'}^{\rm vac} = 2[\theta(-\varepsilon_{N'})-\theta(-\varepsilon_N)] /d_{NN'}$ where $d_{NN'}\coloneqq \varepsilon_{N'}-\varepsilon_N$,
for all Landau levels.
This vacuum subtraction extends the conventional $\tilde B=0$
prescription~\cite{Rischke:2000qz,Rischke:2000ra} to finite magnetic
fields.

After the vacuum subtraction, the normal-state Lindhard contribution,
\(
S_{NN'}^{\rm normal}\coloneqq
\lim_{\Delta\to0}S_{NN'}
=
2[\theta(-\xi_{N'})-\theta(-\xi_N)]/d_{NN'}\,,
\)
vanishes identically in the Meissner response, as required by gauge invariance~%
\footnote{Note that the order of the $\Delta\to0$ and $T\to0$ limits generally matters when checking gauge invariance.}.
We can therefore focus solely on the superconducting contribution
\(
S_{NN'}^{\rm super}\coloneqq
S_{NN'}-S_{NN'}^{\rm normal}
\).
While free from the vacuum divergence, this contribution still
contains the gap-dependent ultraviolet divergence at
$O(|\Delta|^2)$, arising from the extension of the
constant-gap approximation to high Landau levels far from the Fermi
surface.

The remaining ultraviolet contribution is identified as
$S_{NN'}^{\rm pair}
=
\delta_{\lambda,-\lambda'}|\Delta|^2/(2|\varepsilon_N|^3)$.
Evaluating $T_{NN'}$ in the high-energy limit~%
\footnote{The $k$-integrated contribution already closely follows its asymptotic $1/n$ behavior at $n=1$, since the transition matrix element approaches an $O(1)$ function in the high-energy limit.}
and performing the $k$ integral, we find that the contribution from the $n$th Landau level behaves as $M_{{\rm pair},n}^2\propto|\Delta|^2/n$, leading to a logarithmic
divergence in the Landau-level sum.
We remove this divergent contribution by subtracting
$S_{NN'}^{\rm pair}$ for $n\ge n_{\rm F}+1$ up to the cutoff $n_{\rm max}$, where
$n_{\rm F}\coloneqq\lfloor\mu^2/\omega_{\tilde B}^2\rfloor$.
The subtraction involves a finite $O(|\Delta|^2)$ prescription
dependence, which does not affect the leading logarithmic strong-field
asymptotics discussed below~%
\footnote{For $\omega_{\tilde B}\sim\mu$, the high-energy regime
exhibiting the asymptotic $1/n$ behavior may start above
$n_{\rm F}+1$, but shifting the subtraction onset changes only a finite
$O(|\Delta|^2)$ term, which is subleading in this regime.
In the strong-field regime, subtracting only the logarithmic divergence,
\(
\sum_{n=1}^{n_{\rm max}}n^{-1}
=
\log n_{\rm max}+\gamma_E+O(n_{\rm max}^{-1}),
\)
instead of the discrete sum itself leaves a finite
$O(|\Delta|^2)$ difference proportional to $\gamma_E$.}.
We denote the resulting renormalized Meissner mass squared by
$M_{\rm ren}^2$.

\sect{Quantum geometry and topology of the Meissner response}%
Let us now extract the quantum-geometric effect in $S_{NN'}^\text{super}$.
In conventional flat-band superconductors, the quantum-geometric effect is extracted in the isolated-band limit, where the band separation is much greater than the pairing-gap scale, so that the subleading virtual transitions in response functions are negligible \cite{Peotta2015}.
The corresponding limit in the magnetized 2SC is $|d_{NN'}|\gg |\Delta|$, which we call the isolated Landau-level (ILL) limit.

As in the standard argument of the BCS theory, 
we can restrict the longitudinal integral to the vicinity of the Fermi points, $k=\pm k_{\mathrm{F},n}$ with $k_{\mathrm{F},n}\coloneqq\sqrt{\mu^2-n\omega_{\tilde{B}}^2}$.
We define 
the corresponding domain $D_N$
by $|\xi_N|\leq d_\mathrm{F}/2$, where $d_\mathrm{F}\coloneqq \min_{N'} d_{NN'}|_{k=k_{\mathrm{F},n}}= \sqrt{\mu^2+\omega_{\tilde B}^2}-\mu$ so that 
the domains associated with different Landau levels do not overlap.

A systematic expansion of $S_{NN'}$ in powers of $1/d_{NN'}$ then generates the quantum-geometric quantities through $T_{NN'}/(d_{NN'})^2
= a_{NN'}(\delta_{n',n+1} +\delta_{n',n-1})$, 
where the equality follows from $\sigma_i=\rmi[H,x_i]$.
The Meissner mass can then be expressed in terms of the quantum-geometric quantities as
\begin{align}
M_{\mathrm{ILL}}^{2}=
\frac{q_\Delta^2|\Delta|^{2}\omega_{\tilde B}^{2}}{2\pi}
\sum_{n=0}^{
n_\mathrm{F}
}
\int_{D_n^+} \frac{\rmd k}{2\pi}
\frac{\mathcal{G}_n^+}{E_{n}^+}\,,
\label{eq:M_ILL}
\end{align}
where $\mathcal{G}^{+}_n \coloneqq  2(a_{n,n+1}^{++} +  a_{n,n-1}^{++}) + a_{n,n+1}^{+-} + a_{n,n-1}^{+-}$.

In contrast to conventional flat-band superconductors~\cite{Peotta2015}, the present finite pairing-gap contribution is not  
expressed solely in terms of the genuine quantum metric $g^+_n = \sum_{\lambda'}(a_{n,n+1}^{+\lambda'}+a_{n,n-1}^{+\lambda'})$.
This difference arises from 
the particle-antiparticle asymmetry in the pairing gap, $\Delta_+\neq 0$ and $\Delta_-=0$~\footnote{Indeed, if we adopted $\Delta_\pm = \Delta$, Eq.~\eqref{eq:M_ILL} would be expressed in terms of  $2g_n^+$.}.

Although $\mathcal G_n^+$ is not itself the genuine quantum metric, 
the quantum metric provides a lower bound
\( \mathcal G_n^+\geq g_n^+\geq |C|/\omega_{\tilde B}^2 \). Moreover, the LLL obeys the stronger bound \( \mathcal G_0^+\geq 3g_0^+/2 \). We therefore obtain the topological lower bound analogous to that found in Ref.~\cite{Peotta2015},
\begin{equation}
\label{eq:MILL_bound}
 M_{\mathrm{ILL}}^{2} \geq 
 \frac{|C|q_\Delta^2|\Delta|^{2}}{2\pi}\sum_{n=0}^{n_\mathrm{F}}\left(\frac{\delta_{n,0}}{2} + 1 \right) \int_{D_n^+} \frac{\rmd k}{2\pi} \frac{1}{E_{n}^+}\,.
\end{equation}
Therefore, a nonzero Chern number ensures a nonvanishing transverse Meissner response for a finite pairing gap, since the integral on the right-hand side is positive.
Unlike the two-dimensional flat-band setting~\cite{Peotta2015,Liang_2017}, our system retains a finite longitudinal dispersion.
Nevertheless, the Chern-number constraint persists because the topological argument applies only to the transverse sector.

Let us now consider the strong-field limit, $\omega_{\tilde{B}} \gg \mu$. In this limit, only the LLL ($n=0$) contributes, through transitions to the first excited Landau level, and the geometric factor in \eqref{eq:M_ILL} becomes proportional to the LLL quantum metric as $\mathcal G_0^+ \longrightarrow (3/2)g_0^+ = (3/2)|C|/\omega_{\tilde B}^2$.
Consequently, the topological lower bound in Eq.~\eqref{eq:MILL_bound} is saturated, and the leading strong-field behavior is governed by the Chern number. 
Performing the integral analytically,
we obtain~%
\footnote{Finite corrections beyond the ILL approximation can shift
the $O(1)$ constant, including the factor $2$ inside the logarithm,
without changing the leading logarithmic behavior.}
\begin{equation}
\label{eq:M_SF}
M^2_\mathrm{ILL} 
\xrightarrow[\omega_{\tilde B}\gg\mu]{}
\frac{3|C| q_\Delta^2|\Delta|^2}{4\pi^2} \log\frac{2 \omega_{\tilde B}}{|\Delta|} \,,
\end{equation}
which is our main result underlying the scaling highlighted in Eq.~\eqref{eq:M2SC}.
The $|\Delta|^2$ scaling together with
the absence of $\omega_{\tilde B}$ in the prefactor,
originates from the Landau-level quantum geometry dictated by the Chern number $C$.
The logarithm, omitted in Eq.~\eqref{eq:M2SC}, instead originates from the remaining longitudinal dispersion and is thus absent in the fully flat-band limit~\cite{Peotta2015}.
The total $\Delta$ dependence, $\propto|\Delta|^2\log|\Delta|$, is insensitive to the microscopic pairing dynamics, which enters through the value of $\Delta$, as long as the Landau-level structure and a dispersive longitudinal quasiparticle mode are preserved.

\begin{figure}
\centering
\includegraphics[width=0.9\columnwidth]{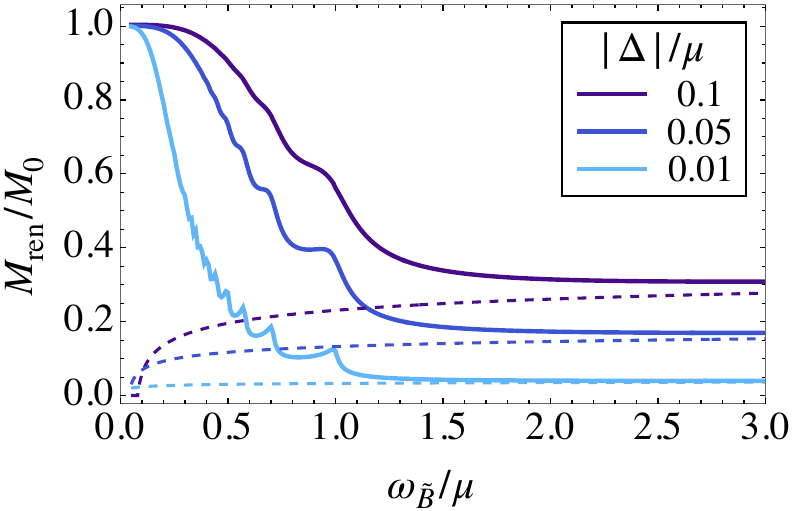}
\caption{
Renormalized transverse Meissner mass normalized by its zero-field value,
as a function of $\omega_{\tilde B}/\mu$ within the constant-gap model.
Solid curves denote the numerical results for
$|\Delta|/\mu=0.1$, $0.05$, and $0.01$.
Dashed curves show 
the $n=0$ contribution.
}
\label{fig:Mren}
\end{figure}

Figure~\ref{fig:Mren} shows the numerical results for the renormalized transverse Meissner mass $M_{\mathrm{ren}}(\tilde{B})$ normalized by its zero-field value $M_0\coloneqq\lim_{\tilde{B}\rightarrow0} M_{\mathrm{ren}}(\tilde{B})$.
Analytically evaluating the Landau-level sum in this limit, we obtain
$M_0\approx \sqrt{2/3}\,q_\Delta\mu/\pi$, in agreement with Ref.~\cite{Rischke:2000qz}.
As the magnetic field increases, the transverse response exhibits a crossover from the conventional regime characterized by the Fermi-surface scale to the quantum-geometric Landau-level regime.
At $\omega_{\tilde B}/\mu\sim1$, the characteristic Fermi-surface contribution of order $\mu$ is strongly suppressed, and the response becomes of order $|\Delta|$, while remaining finite as bounded by Eq.~\eqref{eq:MILL_bound}.
For $\omega_{\tilde B}/\mu > 1$, 
the full numerical results approach the $n=0$ contribution,
demonstrating LLL dominance and the analytic strong-field behavior in Eq.~\eqref{eq:M_SF}.
In the weak-field regime, de Haas--van Alphen oscillations become more pronounced for smaller gaps, reflecting the sharper Fermi surface.

While we have adopted a constant gap for simplicity, the color-superconducting gap generally depends on
momentum~\cite{Son:1998uk,Pisarski:1999tv,Abuki:2001be} and, in a magnetic field, on the Landau-level index~\cite{Yu:2012jn}.
Such dependence modifies quantitative details but not our main result.
For example, adopting the sharp Fermi-surface pairing prescription~\cite{Sen:2015cxa}, \(\Delta_+\to\Delta\theta(n_{\rm F}-n)\) 
reduces the overall coefficient in the strong-field limit Eq.~\eqref{eq:M_SF} to $2/3$ of the constant-gap value~%
\footnote{In the sharp prescription, the gap vanishes for $n_{\rm F}+1$, so that the anomalous $\Delta_+\Delta_+^*$ term in
Eq.~\eqref{eq:S} vanishes for the $n_{\rm F}\to n_{\rm F}+1$
transition. This reduces the coefficient of
$a_{n_{\rm F},n_{\rm F}+1}^{++}$ in $\mathcal G_{n_{\rm F}}^+$ from
two to one. In the strong-field limit, $n_{\rm F}=0$, this gives
$\mathcal G_0^+=g_0^+$ instead of $(3/2)g_0^+$.}, while preserving the $|\Delta|^2\log|\Delta|$ scaling and the characteristic crossover between the conventional and quantum-geometric regimes in
Fig.~\ref{fig:Mren}.
A microscopic determination of $\Delta_n(\tilde B)$ would further refine the quantitative prediction.

\begin{figure}
\centering
\includegraphics[width=0.7\columnwidth]{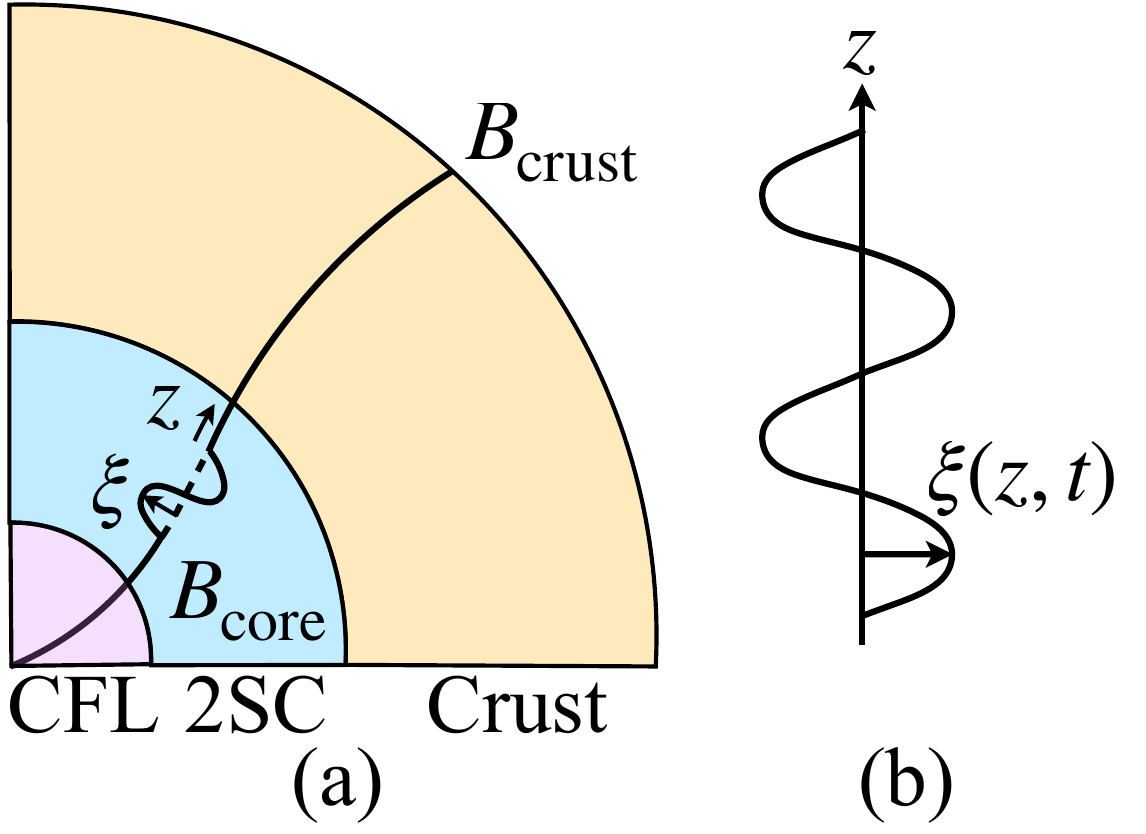}
\caption{
Schematic illustration of magnetic disturbances propagating along a magnetic field line (black curve) in a hybrid magnetar.
(a) A quark core with inner CFL and intermediate 2SC regions surrounded
by a solid crust.
(b) A transverse disturbance $\xi(z,t)$ is modeled as a one-dimensional wave.
}
\label{fig:magnetar}
\end{figure}

\sect{Implications for magnetars}%
The quantum-geometric Meissner response found above may have implications for the kHz QPOs observed in magnetars through its role in the propagation of magnetic disturbances in the quark core.
To explore this possibility, let us consider the model illustrated in Fig.~\ref{fig:magnetar}: a stratified hybrid star with a solid crust and a quark core consisting of an intermediate 2SC layer and an inner CFL region.
Transverse magnetic disturbances $\xi(z,t)$ 
are treated as simplified one-dimensional waves propagating along the magnetic field lines across the crust--2SC and 2SC--CFL interfaces~%
\footnote{At each interface, $\xi$ and the corresponding transverse restoring force $T^\parallel\partial_z\xi$ are continuous, whereas $\partial_z\xi$ can be discontinuous because the effective tension $T^\parallel$ differs between the adjacent regions.}.
The crust is subject to a typical magnetic field $B_\mathrm{crust}\sim 10^{15}\,\mathrm{G}$,
whereas the quark core is assumed to be in the strong-field regime $eB_\mathrm{core} \sim \mu^2$.
For a typical quark chemical potential
$\mu \sim 400\,\mathrm{MeV}$, this condition 
corresponds to a local magnetic field $B_\mathrm{core} \sim 10^{19}\,\mathrm{G}$, which 
remains below theoretical upper bounds estimated from the macroscopic virial limit and local equipartition~\cite{Broderick:2000pe,Ferrer:2010wz,Dexheimer:2011pz}.

In the quark core, the longitudinal flux-tube tension 
scales as $T^\parallel\propto B_\mathrm{core} M^2$~\cite{Alford:2010qf}, where $M$ is the transverse Meissner mass, with  $M_\mathrm{2SC}\sim |\Delta|$ and $M_\mathrm{CFL}\sim \mu$.
The corresponding propagation speeds for magnetic disturbances are $v_\mathrm{2SC} \sim c\sqrt{B_\mathrm{core} }|\Delta|/\mu^2$, and $v_\mathrm{CFL} \sim c\sqrt{B_\mathrm{core}}/\mu$, respectively.
Comparing these with the typical Alfv\'en speed in the crust $v_\text{A}/c \sim 10^{-3}\text{--}10^{-2}$~\footnote{This is estimated from $v_\text{A} \sim B_\mathrm{crust} /\sqrt{4\pi\rho}$ using a typical inner-crust mass density $\rho \sim 10^{12}\text{--}10^{14}\,\mathrm{g/cm^3}$ and $B_\mathrm{crust}\sim 10^{15}\,\mathrm{G}$.},
we find the velocity hierarchy $v_\text{A} \ll v_\mathrm{2SC} \ll v_\mathrm{CFL}$.

Since the wave impedance scales as $Z\sim wv$, where $w$ is the enthalpy density, the velocity hierarchy leads to a strong impedance mismatch, $Z_{\text{A}}\ll Z_\mathrm{2SC}\ll Z_\mathrm{CFL}$, provided variations in $w$ are modest.
Consequently, magnetic disturbances are nearly perfectly reflected at both the 2SC-crust and 2SC-CFL interfaces, allowing the intermediate 2SC region to act as an effective resonant cavity.  
This quantum-geometric cavity strongly suppresses leakage through the interfaces, thereby supporting coherent kHz modes within the 2SC region, in contrast to conventional magneto-elastic scenarios, where oscillations can undergo rapid damping \cite{Levin:2006qd,vanHoven:2011it,Gabler:2012jh,Turolla:2015mwa}.
Nevertheless, the transmission remains finite, allowing a small fraction of the disturbance to leak into the crust along magnetic field lines and act as a source of an observable signature.

The characteristic frequency of magnetic disturbances confined within the 2SC cavity of length $L_{\rm2SC}$ along the magnetic field line is
\begin{align}
f_\mathrm{2SC} \sim \frac{v_\mathrm{2SC}}{L_\mathrm{2SC}} \sim \frac{|\Delta|}{\mu}\frac{c}{L_\mathrm{2SC}}\,.
\end{align}
For typical gap parameters $|\Delta|/\mu \sim 10^{-2}\text{--}10^{-1}$ and a 
characteristic length of $L_{\rm 2SC}\sim1$ to a few km, the characteristic frequency naturally falls in the kHz range. Consequently, our model 
can accommodate the high-frequency QPOs (e.g., $\sim 1.8\,\mathrm{kHz}$) observed in magnetar giant flares~\cite{Israel:2005av,Watts:2006mr,Strohmayer:2006py}.

\sect{Outlook}%
The present work reveals the impact of the quantum metric and the Landau-level topology on Quantum Chromodynamics and its potential astrophysical implications,
complementing recent investigations of the quantum metric in relativistic transport~\cite{Mameda:2025rfn} and Berry monopoles of Weyl fermions in one-flavor color superconductivity~\cite{Sogabe:2024yfl}.
More broadly, quantum-geometric phenomena, actively explored in condensed-matter flat-band systems,  may also emerge in strongly magnetized astrophysical environments, opening new possibilities for observational signatures of color superconductivity.
Furthermore, the Landau-level quantum geometry studied here is not specific to dense quark matter and may also be relevant to other magnetized relativistic systems, including quarks and hadrons in the strong magnetic fields of relativistic heavy-ion collisions.

\begin{acknowledgments}
{\it Acknowledgments}---The authors are grateful to Tetsuo~Hatsuda, Defu~Hou, Kei~Iida, Shuhei~Minato, Dirk~Rischke, Thomas~Sch\"afer, Igor~Shovkovy, Misha~Stephanov, and Yi~Yin for valuable discussions.
N.S. acknowledges the hospitality and financial support of Tokyo University of Science, where part of this work was carried out.
This work was supported by JSPS KAKENHI Grant Numbers 24K17052 (K.\,M.) and 26K17148 (N.\,S.).
\end{acknowledgments}

\bibliographystyle{apsrev4-2}
\bibliography{refs,qm,ckt}

\onecolumngrid

\section{Spectral representation of the transverse Meissner mass}

In the Supplemental Material, we formulate the transverse Meissner response in the full color-flavor space at finite temperature. This provides a unified derivation of the general color–flavor structure and, as a special case, the zero-temperature 2SC result, Eq.~(\ref{eq:M2_bare}).

We start from the inverse Nambu--Gor'kov (NG) propagator in the full color-flavor space,
\begin{align}
\label{eq:Ginv-cf}
\hat{\mathcal G}^{-1}
=
-\partial_\tau
-
(H-\mu)\tau_3
+
\Delta\mathcal M\tau_+
+
\Delta^*\mathcal M^\dagger\tau_-\,,
\end{align}
where
$H=\bm{\sigma}\cdot(-\i\bm\nabla-\widetilde Q\,\widetilde{\bm A}),
$
and
$\mathcal M$ and $\widetilde Q$ denote the gap and rotated-charge matrices, respectively.

Since
$\mathcal M\mathcal M^\dagger$
and
$\widetilde Q$
commute, they can be simultaneously diagonalized.
Introducing the common spectral projectors $\mathcal P_r$,  
\begin{align}
\mathcal M\mathcal M^\dagger
&=
\sum_r
\kappa_r\mathcal P_r\,,
\qquad
\widetilde Q
=
\sum_r
\tilde q_r\mathcal P_r\,,
\end{align}
with $\mathcal P_r\mathcal P_{r'}
=
\delta_{rr'}\mathcal P_r$ and $
\sum_r\mathcal P_r=\bm1$, we decompose 
the full NG propagator into independent eigensectors,
\begin{align}
\hat{\mathcal G}
=
\sum_r
\begin{pmatrix}
\mathcal P_rG_r
&
\widetilde{\mathcal M}_rF_r
\\
\widetilde{\mathcal M}_r^\dagger\bar F_r
&
\mathcal P_r\bar G_r
\end{pmatrix}\,,\qquad \widetilde{\mathcal M}_r=\mathcal M\mathcal P_r\,.
\label{eq:projector-general}
\end{align}
Here, $r$ labels the common eigensectors with eigenvalues $(\kappa_r,\tilde q_r)$.

The polarization tensor in the full color-flavor space is
\begin{align}
\label{eq:K_cal-cf}
\Pi_{ij}^\mathrm{cf}
=2 \cdot \frac{1}{2}
\mathrm{Tr}
\!\left[
\hat{\mathcal G}
Q_X\sigma_i
\hat{\mathcal G}
Q_X\sigma_j
\right]\,,
\end{align}
where $Q_X$ denotes the $\mathrm U(1)_X$ charge matrix, and the factor of 2 corresponds to both chiralities.

Substituting the eigensector decomposition
(\ref{eq:projector-general}) into (\ref{eq:K_cal-cf}) and taking the trace over the Nambu and color-flavor spaces, the transverse Meissner mass squared is given by
\begin{align}
M^2
=2
\sum_{r,r'}
\left[
C_{GG,rr'}
M^2_{GG,rr'}
+
C_{FF,rr'}
M^2_{FF,rr'}
\right]\,,
\label{eq:KGGFF}
\end{align}
where the color-flavor coefficients are
\begin{align}
C_{GG,rr'}
&=
\mathrm{tr}_\mathrm{cf}
\!\left[
Q_X\mathcal{P}_rQ_X\mathcal{P}_{r'}
\right]\,,\qquad
C_{FF,rr'}
=
\mathrm{tr}_\mathrm{cf}
\!\left[
Q_X\widetilde{\mathcal M}_r
Q_X
\widetilde{\mathcal M}_{r'}^\dagger
\right]\,,
\end{align}
while the remaining spin contributions are
\begin{subequations}
\label{eq:KGG,KFF}
\begin{align}
M^2_{GG,rr'}
&=
\frac{T}{2}
\sum_{\omega_\nu}
\frac{1}{2}\sum_{i=x,y}\mathrm{tr}_\sigma
\!\left[
G_r\sigma_iG_{r'}\sigma_i
+
\bar G_r\sigma_i\bar G_{r'}\sigma_i
\right]\,,\\
M^2_{FF,rr'}
&=
-
\frac{T}{2}\sum_{\omega_\nu}
\frac{1}{2}\sum_{i=x,y}
\mathrm{tr}_\sigma
\!\left[
F_r\sigma_i\bar F_{r'}\sigma_i
+
\bar F_r\sigma_iF_{r'}\sigma_i
\right]\,.
\end{align}
\end{subequations}

In the 2SC phase, the gapped sector consists of two common eigensectors,
$r_\pm=(1,\pm\tilde e/2)$,
projected onto by $\mathcal{P}_{r_\pm}$ and corresponding to the paired red-green
$u$- and $d$-quark subspaces, respectively.
Both satisfy
$\mathrm{tr}\,\mathcal{P}_{r_\pm}=2$,
reflecting the residual
$\mathrm{SU}(2)_\mathrm{c}$
color degeneracy.
In the weak photon-gluon mixing limit ($e\ll g$),
$\tilde e\simeq e$,
$e_X\simeq g$,
and
$Q_X\simeq e_X T_8$.
Since
$T_8$
acts as
$\bm1/(2\sqrt3)$
on the paired red-green subspace,
$
Q_X\mathcal{P}_{r_\pm}=q_X\mathcal{P}_{r_\pm}=e_X\mathcal{P}_{r_\pm}/(2\sqrt3)
$
and
$
Q_X\mathcal M=e_X\mathcal M/(2\sqrt3).
$
Therefore,
\begin{align}
\label{eq:C-2SC}
C_{GG,r_\pm r_\pm}
=
C_{FF,r_\pm r_\pm}
\simeq \frac{q_\Delta^2}{2}\,,
\end{align}
with $q_\Delta=2 q_X$. 
The off-diagonal coefficients vanish,
e.g.,
$
C_{GG,r_+r_-}=0,
$
so that
\begin{align}
C_{GG,rr'}
=
\delta_{rr'}C_{GG,r}\,,
\qquad
C_{FF,rr'}
=
\delta_{rr'}C_{FF,r}.
\end{align}
Thus,
\begin{align}
\label{eq:Msum}
M^2=\sum_r M_r^2\,.
\end{align}
Projecting the general color-flavor formulation onto a paired eigensector yields the NG propagator and the current-current correlation function used in the main text.

In each eigensector, the block propagators (\ref{eq:projector-general}) admit the following Landau-level spectral representation:
\begin{align}
G_r(\i\omega_\nu)
&=
-
\sum_{\lambda=\pm}
\sum_{n,\ell}
\int \frac{\rmd k}{2\pi}
\frac{\i\omega_\nu+\xi_N
}{
\omega_\nu^2+E_N^2
}
|u_N\rangle
\langle u_N|\,,\quad
F_r(\i\omega_\nu)
=
\sum_{\lambda=\pm}
\sum_{n,\ell}
\int \frac{\rmd k}{2\pi}
\frac{\Delta_\lambda}
{\omega_\nu^2+E_N^2}
|u_N\rangle
\langle u_N|\,.
\label{eq:GrFr}
\end{align}
Here, $\bar G_r$ and $\bar F_r$ are obtained by the substitutions
$
\xi_N\rightarrow-\xi_N
$
and
$
\Delta_\lambda\rightarrow\Delta_\lambda^*.
$
The Landau-level eigenstates $|u_N\rangle$ given by (\ref{eq:eigen_state}) are independent of the guiding-center degeneracy label $\ell$, which therefore contributes only through the overall Landau-level degeneracy.

Substituting Eq.~(\ref{eq:GrFr}) into Eq.~(\ref{eq:KGG,KFF}) and
performing the Matsubara sum, we obtain
\begin{align}
M_r^2(\tilde B,T)
&=
\frac{q_\Delta^2\omega_{\tilde B}^2}{8\pi}
\sum_{n,n'}
\sum_{\lambda,\lambda'}
\int \frac{\rmd k}{2\pi}
T_{NN'}(k)\,
S_{NN'}(k,T)\,.
\end{align}
Here, $T_{NN'}$ is given by Eq.~\eqref{eq:TNN'-def}, while $S_{NN'}(k,T)$ is the finite-temperature generalization of Eq.~\eqref{eq:S}, given by
\begin{align}
S_{NN'}(k,T)
&= 
\left[
1+\frac{\xi_N\xi_{N'}}{E_NE_{N'}}
+
\frac{\Delta_\lambda\Delta_{\lambda'}^*}{E_NE_{N'}}
\right]
R_{NN'}
-
\left[
1-\frac{\xi_N\xi_{N'}}{E_NE_{N'}}
-
\frac{\Delta_\lambda\Delta_{\lambda'}^*}{E_NE_{N'}}
\right]
\Phi_{NN'}\,,
\end{align}
where we have used \eqref{eq:C-2SC} and defined
\begin{align}
R_{NN'}
&=
\frac{
f_\mathrm{F}(E_N)-f_\mathrm{F}(E_{N'})
}{
E_N-E_{N'}
},
\qquad
\Phi_{NN'}
=
\frac{
1-f_\mathrm{F}(E_N)-f_\mathrm{F}(E_{N'})
}{
E_N+E_{N'}
}\,,
\end{align}
with $f_\mathrm{F}(E)= (\rme^{E/T}+1)^{-1}$.

At zero temperature,
$
f_\mathrm{F}(E_N)\rightarrow0,
$
so that
$
R_{n,n+1}^{\lambda\lambda'}\rightarrow0
$
and
$
\Phi_{n,n+1}^{\lambda\lambda'}
\rightarrow
(E_N+E_{n+1}^{\lambda'})^{-1}.
$
In the 2SC phase, the summation over $r$ in Eq.~(\ref{eq:Msum}) runs over the two paired eigensectors,
$r=r_\pm$.
Since the corresponding color-flavor coefficients, Eq.~(\ref{eq:C-2SC}), are identical, the two sectors give equal contributions.
We thus recover Eq.~(\ref{eq:M2_bare}).

\end{document}